\documentclass[
 reprint,
superscriptaddress,
 amsmath,amssymb,
 aps, physrev,
]{revtex4-2}
\usepackage[version=3]{mhchem} 
\usepackage{float}
\usepackage{comment}
\usepackage{esint}
\usepackage{tabularx}
\usepackage{soul}
\usepackage{amsmath}
\usepackage{amssymb}
\usepackage[normalem]{ulem}

\usepackage[utf8]{inputenc}
\usepackage[T2A,T1]{fontenc}
\usepackage[english]{babel}
\usepackage[
colorlinks=true,bookmarks=false,citecolor=blue,urlcolor=blue
]
{hyperref}

\usepackage{graphicx}
\usepackage{color}
\usepackage{soul}
\usepackage{xcolor}
\usepackage{comment}

\usepackage{xfrac}
\usepackage{gensymb}
\usepackage[caption=false]{subfig}

\usepackage{dcolumn}
\usepackage{bm}

\newcommand{\pvec}[1]{\vec{#1}\mkern2mu\vphantom{#1}}

\renewcommand{\bm}{\mathbf{m} } 

\renewcommand{\Im}{\frak{I}\mathrm{m} }

\newcommand{\eps}{\varepsilon}

\begin{document}
\preprint{APS/123-QED}
\title{\textbf{Strong coupling of chiral light with chiral matter: a macroscopic study}} 
\author{Sergey A. Dyakov}
\affiliation{Skolkovo Institute of Science and Technology, Moscow, Russia}

\author{Ilia A. Smagin}
\affiliation{Skolkovo Institute of Science and Technology, Moscow, Russia}

\author{Natalia S. Salakhova}
\affiliation{Skolkovo Institute of Science and Technology, Moscow, Russia}

\author{Oleg Blokhin}
\affiliation{Moscow Center for Advanced Studies, Moscow, 123592, Russia}

\author{Denis G. Baranov}
\affiliation{Moscow Center for Advanced Studies, Moscow, 123592, Russia}

\author{Ilia M. Fradkin}
\affiliation{Skolkovo Institute of Science and Technology, Moscow, Russia}
\affiliation{Moscow Center for Advanced Studies, Moscow, 123592, Russia}

\author{Nikolay A. Gippius}
\affiliation{Skolkovo Institute of Science and Technology, Moscow, Russia}

\date{\today}
\begin{abstract}
Maximizing the interaction between chiral light and chiral matter is pivotal for the advancement of technologies enabling optical detection that distinguishes between different handedness in chiral organic molecules. One strategy involves developing a resonator that sustains photonic modes with non-zero electromagnetic handedness, which interact differently with chiral molecules of opposite enantiomers. When chiral molecules are positioned in resonator hotspots, they can alter the system's characteristics due to their inherent electric and magnetic transition dipole moments. In this study, we explore this interaction by incorporating the Lorentz pole into the macroscopic parameters of the chiral medium: dielectric permittivity, magnetic permeability, and chirality coefficient. The latter, also known as the Pasteur parameter, is a dimensionless macroscopic measure indicating the medium's chirality, interlinking electric and magnetic fields in the constitutive relations. We show that introducing the Lorentz pole into these macroscopic material parameters of the chiral medium results in chiral strong coupling between light and matter, with the strength of coupling determined by both the medium's chirality and the photonic mode's chirality.
\end{abstract}

\maketitle
\newpage


\section*{Introduction}
Chirality is a key phenomenon that describes an object's ability to have two non-superimposable mirror-image forms, known as enantiomers. This concept was scientifically defined over a century ago \cite{kelvin1894molecular} and has gained significant attention in recent years due to its vital role in natural phenomena. Chirality is crucial across multiple disciplines such as chemistry, biology, and physics, affecting molecular behavior, material properties, and light-matter interactions. Chiral matter can be composed of either a) chiral molecules, which cannot match their mirror image, such as amino acids and sugars \cite{nguyen2006chiral, Deutsche1969}, or b) achiral molecules arranged in chiral patterns, such as in cholesteric liquid crystals \cite{Panov2003,Panov2006} \footnote{In fact, liquid crystal molecules are often chiral, but configurational chirality can greatly surpass constitutional chirality in optical effects when the structural periodicity is near the wavelength in use.}. On the other hand, a basic example of chiral light is a circularly polarized plane wave, which can either be right-handed -- exhibiting counterclockwise rotation of the electric field vector when looking in the direction of wave propagation -- or left-handed, exhibiting clockwise rotation. Other electromagnetic fields, like vortex beams or interfering circularly polarized waves, can also be chiral. Their chirality can be quantified using chirality density and chirality flow density metrics \cite{bliokh2011characterizing}. 

The interaction between chiral light and chiral matter unveils numerous intriguing optical phenomena, including circular birefringence, circular dichroism, and chiral symmetry breaking \cite{barron2009molecular, inoue2004chiral}. Circular dichroism describes the difference in extinction of left- and right-handed circularly polarized light by chiral molecules, causing a variance in the intensity of light that is either transmitted or reflected. Circular dichroism spectroscopy is an effective method for examining the structure and conformation changes in chiral molecules \cite{Johnson1988, Hendry2010, Tang2011}. Circular birefringence refers to how chiral substances transmit left- and right-handed circularly polarized light at different phase velocities, leading to the rotation of the polarization plane of linearly polarized light as it travels through. The extent and direction of optical rotation are influenced by the molecular constitution and concentration of chiral molecules within the material \cite{inoue2004chiral, hodgkinson2001inorganic}. Finally, chiral symmetry breaking occurs when chiral light encourages the dominant formation of one chiral variant over another. This phenomenon holds significant relevance in asymmetric synthesis and chiral catalysis, affecting chemical reaction outcomes \cite{McConnell2007}. All the above effects are the manifestations of optical activity, which occurs only in chiral materials.

The investigation of optical activity is crucial for grasping the basic concepts of light-matter interactions, with applications spanning fields such as chemistry, biology, materials science, and optics \cite{tang2010optical, dyakov2018magnetic, genet2022chiral, fradkin2022plasmonic, baranov2023toward, lodahl2017chiral}. It has the potential to provide innovative techniques for manipulating the properties of chiral molecules and materials, which could culminate in breakthroughs in drug development, chiral molecule detection, and the creation of advanced optical devices \cite{Johnson1988, Tang2011, Shapiro2003, McConnell2007,Ma2013,dyakov2020vertical,chambers1999stratified, fradkin2023nearly,Baranov2019,petersen2014chiral}. The success of these promising developments relies on our understanding of chiral light-matter interaction regimes. In Refs.~\cite{schafer2023chiral, Riso2023} a chiral light-matter strong coupling has been theoretically predicted using nonperturbative analytical models of an ensemble of chiral molecules with a chiral optical mode. Since chiral strong coupling can induce enantiospecific optical response, creating cost-effective and efficient resonators supporting chiral electromagnetic eigenmodes becomes a cumbersome stone in chiral polaritons.  

\begin{figure}[t!]
\centering\includegraphics[width=0.8\columnwidth]{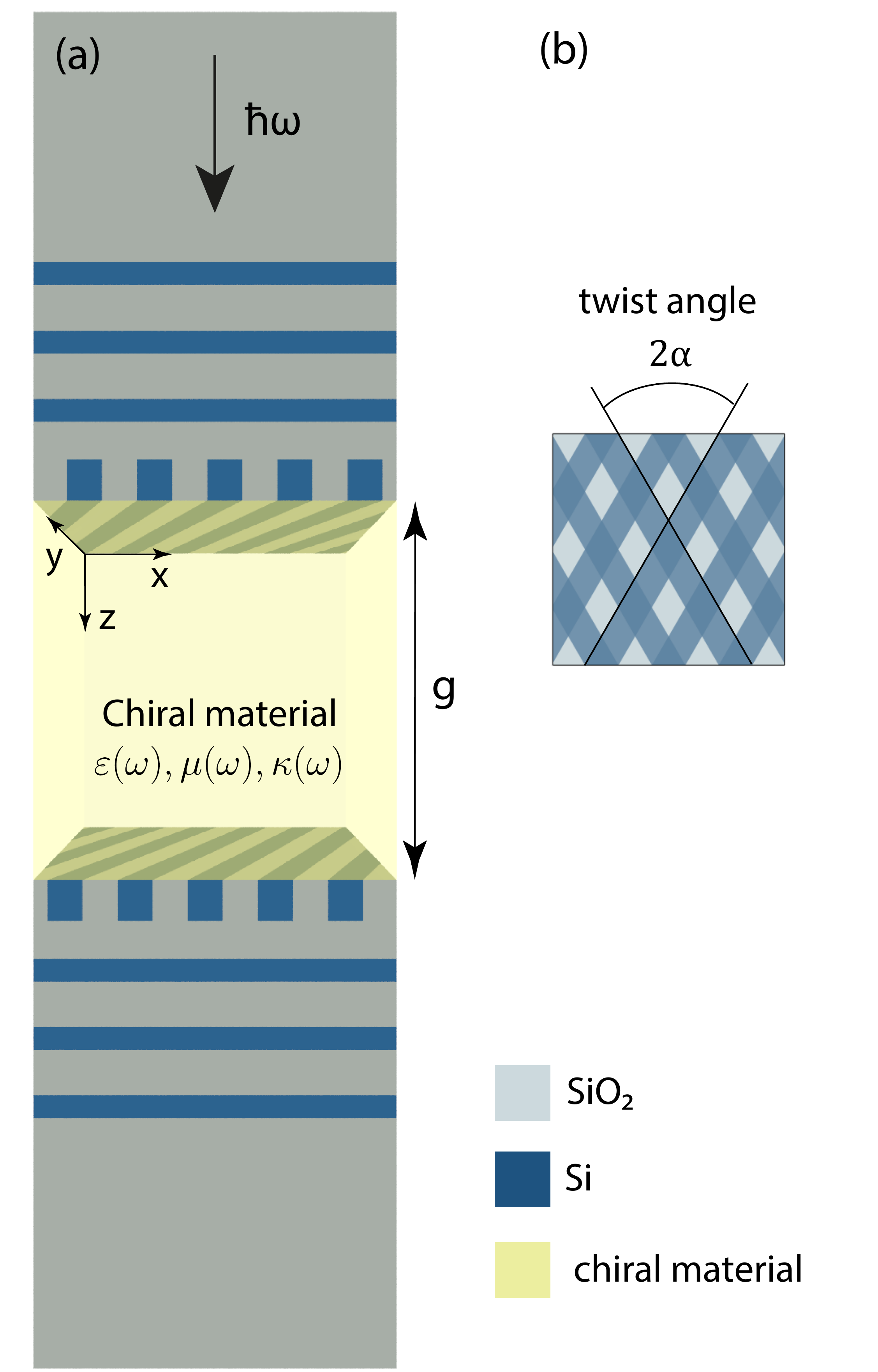}
\caption{(a) The sketch of a chiral Fabry-Pérot resonator comprising two anisotropic handedness-preserving mirrors. Each handedness-preserving mirror consists of a one-dimensional silicon grating on top of the Bragg mirror on a silica substrate. (b) Orientations of gratings.}
\label{fig:sample}
\end{figure}

Resonators supporting chiral modes can be created with resonant particles and metasurfaces. It has been demonstrated in the literature \cite{Tang2011, Mohammadi2019, Wu2013, Graf2019, dyakov2021photonic, yoo2015chiral, petersen2014chiral, shomroni2014all} that even structures lacking inherent chirality can sustain chiral electromagnetic fields. However, in such scenarios, the effect is localized, resulting in a low volume-averaged chirality density. An alternative method involves utilizing spiral or helically designed metasurfaces \cite{Kan2015, FernandezCorbaton2019, Hendry2010, Pham2016, Wang2023, Konishi2011, Barbillon2020, shi2022planar, wang2022metasurface}. Recently, a plasmonic metasurface containing spiral-like elements has been leveraged to achieve a strong coupling regime between chiral electromagnetic modes and chiral molecules in their angular-resolved photoluminescence response \cite{Kumar2024}. As a result, an experimentally notable difference in the photoluminescence lifetime has been demonstrated. While metasurface structures can effectively support chiral eigenmodes, the spatial confinement of chiral light remains restricted to a two-dimensional surface. 

With that in mind, Fabry-Pérot resonators that can sustain modes with specific handedness throughout the entire cavity volume present an intriguing possibility. Unfortunately, forming such Fabry-Pérot resonators using two homogeneous isotropic plates is impossible, since upon reflection such highly-symmetrical plates change the handedness of light to the opposite. To reduce the system's symmetry, $C_{6v}$-symmetric photonic crystal slabs, along with a high in-plane wavevector for eigenmodes, can be employed \cite{feis2020helicity}. In this configuration, first-order diffraction facilitates upward and downward propagation of modes with the same handedness, thereby substantially increasing the volume of the chiral field compared to localized modes. Nevertheless, an experimentally preferable setup for analyzing chiral molecules remains with normal light incidence; hence, the need persists for a Fabry-Pérot resonator capable of supporting normally propagating modes with identical handedness.

Fabry-Pérot resonators supporting chiral normally propagating modes can be designed by breaking the spatial inversion or the time-reversal symmetry \cite{Hubener2021}. A resonator with a broken spatial inversion can be formed by a pair of handedness-preserving mirrors, reflecting circularly-polarized light into light with the same handedness. Within this resonator, a helical mode may exist with the electric field tracing a helical path through space while maintaining local linear polarization at every coordinate \cite{Hubener2021, Plum2015, dyakov2024chiral}. In a Fabry-Pérot resonator formed by Faraday mirrors, time-reversal symmetry is broken; as a result, such a structure can support a chiral Faraday mode, where two standing waves of opposite circular polarization form with an alternating nodal structure \cite{Hubener2021}. In this work, we are focused on structures with spatial inversion breaking.

In Ref.\,\cite{Plum2015} handedness-preserving mirrors based on a 2D array of split metallic rings have been demonstrated. This object, characterized by its low symmetry, showcases intrinsic chirality in its structural components, facilitating different interactions with left- and right-handed circularly polarized light, ultimately maintaining the polarization state after reflection. More recently, B.~Semnani et al. \cite{Semnani2020} suggested and verified a mirror for near-infrared wavelengths, employing a photonic crystal slab with a chiral multi-hole pattern in each unit cell. This mirror maintained the light's handedness by carefully manipulating the hybridization of the TE and TM modes within the 2D photonic crystal. Furthermore, various adaptations of handedness-preserving mirrors across multiple frequency ranges and material bases have been demonstrated in Refs.~\cite{li2020spin, Liu2020, gorkunov2021bound, Voronin2022}. It has been illustrated that Fabry-Pérot cavities, constructed with such mirrors, can locally enhance one handedness over another, or even generate standing waves with a specific handedness \cite{feis2020helicity, Sun2022, gautier2022planar, mauro2023chiral, Voronin2022}. The second scenario is particularly compelling because it enables the formation of resonant modes free from antinodes, featuring high chirality density across the entire cavity gap.

Recently, in Ref.~\cite{dyakov2024chiral}, a Fabry-Pérot resonator consisting of a pair of twisted anisotropic handedness-preserving mirrors was demonstrated. The preservation of the handedness of the incoming light by these mirrors is governed by the fulfillment of a condition where the amplitude reflection coefficients for two orthogonal polarizations have a phase difference of $\pi$. The resonant gap size of this system can be mechanically controlled by varying the angle between the anisotropy axes of the mirrors. Due to the uniform distribution of field intensity and chirality density of its modes, such a Fabry-Pérot resonator can operate as a platform for the enhancement of chiral light-matter interaction.

In the literature, the effect of the chirality of matter on the properties of a resonator is often omitted. Nevertheless, in some cases, it is implied that the chiral medium is present in the system in some form. In numerical solvers, chiral matter can be represented by small chiral particles of helix-like shapes \cite{schaferling2012tailoring, schaferling2014helical, liu2009stereometamaterials, kuzyk2012dna, avalos2022chiral}, chiral dipoles \cite{Voronin2022, dyakov2024chiral} or inherently chiral medium consisting of chiral molecules. The latter requires constitutive relations for Maxwell's equations in a special form, such that they could describe optical activity in chiral media. One of the simplest and most convenient constitutive relations was phenomenologically introduced in \cite{Fedorov} by Fedorov et al. in 1976; they include local macroscopic parameters of the medium connecting the electric induction vector $\vec{D}$ with the magnetic field $\vec{H}$, and also magnetic induction $\vec{B}$ with the electric vector $\vec{E}$. Accurate account for the presence of a chiral medium in the system should incorporate these relations. It should be noted that there are alternative ways of accounting for the chirality of the medium \cite{Fedorov}.

In this work, we aim at theoretical and numerical studies of the interaction of chiral electromagnetic modes with chiral matter in terms of macroscopic electrodynamics. We will design a Fabry-Pérot resonator supporting chiral modes based on anisotropic wide-band handedness-preserving mirrors. Next, in the gap between the mirrors, we will place a chiral medium characterized by a macroscopic chirality parameter, analogous to macroscopic dielectric permittivity and magnetic permeability. To properly capture the interaction of electromagnetic modes of the chiral Fabry-Pérot resonator with a resonant chiral medium, we will describe the response of the medium with the chiral Lorentz model. We will demonstrate that the presence of the material resonance in the macroscopic parameters of the medium results in strong coupling between chiral photonic modes and the chiral medium. The strength of this interaction is determined by the compatibility of the enantiomers of the mode and the medium.

\section{Theory}
\label{sec:theory}
\subsection{Macroscopic description of chiral medium}  
Extensive research of optical activity in the past century \cite{fyodorov1976teoriya, lindell1994electromagnetic, simovski2018composite} can be summarized in the following constitutive relations, which are applicable to a general bi-anisotropic medium:
\begin{align}
\begin{split}
    \vec{D} &= \hat{\varepsilon}(\omega)\vec{E}+ \hat{\eta}(\omega)\vec{H}\\
    \vec{B} &= \hat{\zeta}(\omega)\vec{E}+ \hat{\mu}(\omega)\vec{H},
\end{split}
\end{align}
where "$\wedge$" symbol denotes tensor functions which represent local macroscopic parameters of the medium. These relations are Lorentz invariant and describe all known types of gyrotropy \cite{fyodorov1976teoriya}. In this work, we focus on a bi-isotropic medium and replace tensors with scalars. It can be shown that for a reciprocal medium the condition $\eta = -\zeta$ must be satisfied \cite{lindell1994electromagnetic}. Because of that, it is advantageous to replace the parameters $\eta$ and $\zeta$ by
\begin{align}
\label{eq:new}
\begin{split}
    \eta &= \chi+i\kappa\\
    \zeta &= \chi-i\kappa,     
\end{split}
\end{align}
where $\kappa$ and $\chi$ are complex numbers and are referred to as macroscopic chirality and non-reciprocity parameters of the medium \cite{lindell1994electromagnetic}, also known as Pasteur parameter and Tellegen parameter, respectively. It can be demonstrated that relative macroscopic chirality and non-reciprocity parameters can also be introduced
\begin{equation}
\label{eq:rel}
    \kappa_r \equiv \frac{\kappa}{\sqrt{\varepsilon\mu}}, ~~~ \chi_r \equiv \frac{\chi}{\sqrt{\varepsilon\mu}}.
\end{equation}
Considering only reciprocal media, we will use constitutive relations in the following form:
\begin{align}
\begin{split}
    \vec{D} &= \phantom{-i}\varepsilon(\omega)\vec{E}+ i\kappa(\omega)\vec{H}\\
    \vec{B} &= -i\kappa(\omega)\vec{E}+ \mu(\omega)\vec{H}.    
\end{split}
\end{align}
Based on these relations, it can be shown that in a chiral medium, the refractive indices for the left and right circularly polarized waves are different and are expressed as:
\begin{equation}
    (n^{\mathrm{LCP,~RCP}})^2 = (\sqrt{\varepsilon\mu} \pm \kappa)^2.
\end{equation}
This expression can be recast in a more profound form
\begin{equation}
    \label{eq:ncp}
    n(\Lambda) = \sqrt{\varepsilon\mu} + \Lambda \kappa,
\end{equation}
where $\Lambda = \pm 1$ are the eigenvalues of the helicity operator \cite{FernandezCorbaton2016} 
\begin{equation}
    \hat\Lambda = 
    \frac{1}{k}\begin{pmatrix}
        \nabla\times & 0\\0 &\nabla\times
    \end{pmatrix},
\end{equation}
acting on the complex hypervector of the electromagnetic field $\left[\vec{E},Z\vec{H} \right]$, where $Z = \sqrt{\mu/\varepsilon}$ is the medium impedance and $k=\omega/c$ is the magnitude of the wavevector. The corresponding eigenvectors represent circularly polarized waves
\begin{equation}
    Z\vec{H} = - i \Lambda\vec{E},
\end{equation}
with $\Lambda=+1$ and $\Lambda = -1$ describing, by definition, LH (left-circularly polarized) and RH (right-circularly polarized) waves, respectively.

From formula \eqref{eq:ncp}, it follows that the real part of the chirality parameter $\kappa$ is responsible for different refractive indices of LH and RH plane waves, while the imaginary part of $\kappa$ induces their different absorption. Obviously, the difference in the refractive indices of the LH and RH waves causes optical rotation of the polarization direction of a linearly polarized plane wave when it propagates in a chiral medium. 

\subsection{Fabry-Pérot cavity with chiral medium}
\label{sec:theory:FP}
As indicated in the introduction, enhancement of chiral light-matter interaction can be realized by designing a Fabry-Pérot resonator that supports chiral electromagnetic modes and placing a chiral medium in it. Such a resonator can be formed by a pair of handedness-preserving mirrors which reflect left-handed (LH) and right-handed (RH) incident plane waves into the waves of the same handedness. This feature can be achieved in an anisotropic mirror if it satisfies the condition $r_{xx}=-r_{yy}$, where $r_{xx}$ and $r_{yy}$ are co-polarized amplitude reflection coefficients \cite{dyakov2024chiral}. A Fabry-Pérot resonator comprised of these mirrors supports chiral electromagnetic modes if the in-plane optical axes of the upper and lower mirrors are not parallel or perpendicular to each other but form a certain angle \cite{dyakov2024chiral}, as shown in Fig.~\ref{fig:sample}. This configuration can be obtained by rotation of the upper and lower mirrors in the horizontal plane relative to their initial orientation along the $y$-axis.

To characterize the eigenmodes of such a Fabry-Pérot resonator, we employ a method that represents the local field by a complex vector of amplitudes \cite{menzel2010advanced, dyakov2024chiral}. The vector of amplitudes can be expressed either in the linear polarization basis (indicated as "$xy$") or in the circular polarization basis (indicated as "$\rho\sigma$"):
\begin{equation}
    \pvec{\mathrm{A}}_{xy} = 
    \begin{bmatrix}
        a_x\\a_y
    \end{bmatrix},~~~~~~~
    \pvec{\mathrm{A}}_{\rho\sigma} = 
    \begin{bmatrix}
        a_\rho\\a_\sigma
    \end{bmatrix},
\end{equation}
with an associated transformation matrix $\mathbb{T}$:
\begin{equation}
    \mathbb{T}\equiv\mathbb{T}_{\rho\sigma \leftarrow xy} = 
    \frac{1}{\sqrt{2}}
    \begin{bmatrix}
        1 & i \\ 1 & -i
    \end{bmatrix},
\end{equation}
such that $\pvec{\mathrm{A}}_{\rho\sigma} = \mathbb{T}\pvec{\mathrm{A}}_{xy}$. Given that the vector of amplitudes describes waves traveling in either the positive or negative $z$-direction, we use the notation $\pvec{\mathrm{A}}^+$ for waves propagating positively and $\pvec{\mathrm{A}}^-$ for those propagating negatively. The vectors $\pvec{\mathrm{A}}^+_{\rho\sigma} = [1, 0]^T$ and $\pvec{\mathrm{A}}^-_{\rho\sigma} = [0, 1]^T$ represent the RH wave, whereas the vectors $\pvec{\mathrm{A}}^-_{\rho\sigma} = [1, 0]^T$ and $\pvec{\mathrm{A}}^+_{\rho\sigma} = [0, 1]^T$ describe the LH wave.

Without loss of generality, we assume that the vectors $\pvec{\mathrm{A}}^\pm$ are considered near the lower mirror. Under this assumption, the eigenmodes of the Fabry-Pérot resonator are obtained by solving the following eigenvalue equation:
\begin{equation}
    \label{eigenproblem}
    \mathbb{M}\pvec{\mathrm{A}}^+\equiv \mathbb{P}^{+}\mathbb{R}^\mathrm{upper}\mathbb{P}^{-}\mathbb{R}^\mathrm{lower}\pvec{\mathrm{A}}^+ = m\pvec{\mathrm{A}}^+,
\end{equation}
where the eigenvalue $m$ is set to 1. In Eqn.~\eqref{eigenproblem}, $\mathbb{P}^{\pm}$ are the matrices accounting for wave propagation in the positive and negative $z$ directions, and $\mathbb{R}^\mathrm{upper}$, $\mathbb{R}^\mathrm{lower}$ are the reflection matrices for the rotated upper and lower mirrors, respectively.

In a chiral medium, the propagation matrices $\mathbb{P}^{+}$ and $\mathbb{P}^{-}$ differ because LH and RH polarized waves have distinct phase velocities. Within the $\rho\sigma$-basis, the propagation matrices are represented as follows:
\begin{equation}
    \mathbb{P}^{\pm} = 
    \begin{bmatrix}
        e^{ikg(\sqrt{\varepsilon\mu}\mp \kappa)} & 0\\ 0 & e^{ikg(\sqrt{\varepsilon\mu}\pm \kappa)}
    \end{bmatrix},
\end{equation}
where $g$ denotes the gap between the mirrors. 

Next, the reflection matrices $\mathbb{R}^\mathrm{upper}$ and $\mathbb{R}^\mathrm{lower}$ of the rotated upper and lower mirrors can be obtained from that of an unrotated mirror by applying the rotation operator. For an optimized handedness-preserving mirror with the in-plane anisotropy axis initially oriented along the $y$-axis, the initial reflection matrix has the following form in the $\rho\sigma$-basis:
\begin{equation}
    \label{form15}
    \mathbb{R}_{\rho\sigma}^0 =  
    \begin{bmatrix} 
    0 & r \\ r & 0
    \end{bmatrix},
\end{equation}
that corresponds to the following reflection matrix in the $xy$-basis:
\begin{equation}
    \mathbb{R}_{xy}^0 =  
    \begin{bmatrix} 
    r & 0 \\ 0 & -r
    \end{bmatrix}.
\end{equation}
For simplicity, we also assume the amplitude reflection coefficient is fixed such that the equality $|r|=1$ holds for the entire frequency range of interest. Then, to make the system chiral, we rotate the upper and lower anisotropic mirrors about the $z$-axis by angles $-\alpha$ and $\alpha$ respectively; as a result, the upper and lower reflection matrices in such a system are expressed as
\begin{equation}
    {\mathbb{R}}_{xy}^\mathrm{upper} = \mathbb{S}^{-1}\mathbb{R}^0_{xy}\mathbb{S}, ~~~~~  {\mathbb{R}}_{xy}^\mathrm{lower} = \mathbb{S}\mathbb{R}^0_{xy}\mathbb{S}^{-1},
\end{equation}
where $\mathbb{S}$ is a rotation matrix:
\begin{equation}
    \mathbb{S} = \begin{bmatrix}
        \cos{\alpha} & -\sin{\alpha}\\\sin{\alpha}& \cos{\alpha}
    \end{bmatrix}.
\end{equation}
After some algebra, we obtain the following expression for the matrix $\mathbb{M}$ in circular basis:
\begin{equation}
\label{eq:M}
    \mathbb{M}_{\rho\sigma} = 
    \begin{bmatrix}
        r^2e^{2ikg(\sqrt{\varepsilon\mu} - \kappa)}e^{-4i\alpha} & 0\\ 0& r^2e^{2ikg(\sqrt{\varepsilon\mu} + \kappa)}e^{+4i\alpha}
    \end{bmatrix}.
\end{equation}
The corresponding eigenvalue problem has two series of non-trivial solutions described by the following basis vectors and eigenvalues for RH and LH waves:
\begin{align}
    \label{eq:Am}
    \begin{split}
    \pvec{\mathrm{A}}_{\rho\sigma}^{+(1)} &= \begin{bmatrix} 1\\0 \end{bmatrix},~~~~~~ m^{(1)}=r^2e^{2ikg(\sqrt{\varepsilon\mu} - \kappa)}e^{-4i\alpha}\\
    \pvec{\mathrm{A}}_{\rho\sigma}^{+(2)} &= \begin{bmatrix} 0\\1 \end{bmatrix},~~~~~~ m^{(2)}=r^2e^{2ikg(\sqrt{\varepsilon\mu} + \kappa)}e^{+4i\alpha},
    \end{split}
\end{align} which at $m^{(1,2)} = 1$ give the resonant frequencies of the Fabry-Pérot resonator:
\begin{equation}
    \label{eq:finalg}
    \frac{\omega}{c} = \frac{-\mathrm{arg} ~r \pm 2\alpha + \pi N}{g(\sqrt{\varepsilon\mu} \mp \kappa)},
\end{equation}
where signs "-" and "+" in the enumerator correspond to the first and second solutions respectively, and $N$ is an integer.

According to the definitions given above for the mode's handedness in terms of the vector of amplitudes, the first solution corresponds to the RH mode, while the second one corresponds to the LH mode. Note that in a lossless medium with positive $\kappa$, the refractive index of the LH wave is larger than that of the RH wave. In a lossless medium with negative $\kappa$, the opposite is true.

As one can see from \eqref{eq:Am}, if the medium between the mirrors is non-chiral ($\kappa=0$), the eigenvalues of matrix \eqref{eq:M} are degenerate at $\alpha=0$, $\pi/4$, and $\pi/2$. This is because at these angles the system acquires an additional symmetry plane becoming non-chiral. At other angles $\alpha$, the degeneracy is lifted and the structure is geometrically chiral, supporting either RH or LH mode. If the right-hand side of Eqn.~\eqref{eq:finalg} depends on $\omega$ (which is true for dispersive materials), the analytical solution can only be found for special cases. Note that in the case of non-chiral material ($\kappa = 0$), the system has two double-degenerate modes, with equal phase velocities of positively and negatively propagating circularly-polarized waves \cite{dyakov2024chiral}.

\begin{figure*}[t!]
\centering\includegraphics[width=1\textwidth]{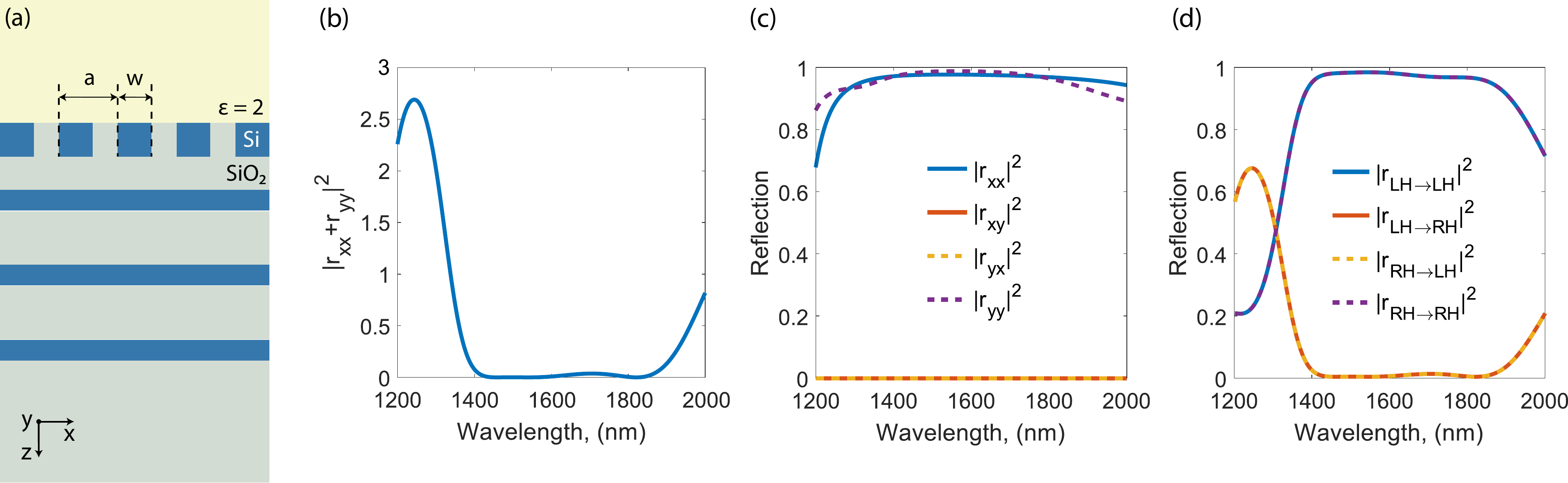}
\caption{(a) The sketch of a wide-band handedness-preserving mirror consisting of the one-dimensional array of Si stripes on a Si/SiO$_2$ multilayered structure. Thicknesses of the layers are the following: 636~nm (grating), 112~nm, 114~nm, 276~nm, 114~nm, 276~nm, 114~nm. The period $a = 450$~nm, the stripes' width $w = 90$~nm. (b) Spectral dependence of $|r_{xx}+r_{yy}|^2$. (c) Spectral dependencies of cross-polarization reflection coefficients in the basis of linear polarizations. (d) Spectral dependencies of cross-polarization reflection coefficients in the basis of circular polarizations.}
\label{fig:mirror}
\end{figure*}

To properly capture the interaction of electromagnetic modes of the chiral Fabry-Pérot resonator with a resonant chiral medium, in the vicinity of the material resonance, we describe the response of the medium with the chiral Lorentz model \cite{Condon1937, baranov2023toward}:
\begin{align}   
\label{eq:lorenz}
\begin{split}
    \varepsilon(\omega) & = \varepsilon_{\infty} + f_e\frac{\omega_0^2}{\omega_0^2 - \omega^2 - i \gamma \omega}, \\
    \kappa(\omega) &= f_{me} \frac{\omega}{\omega_0} \cdot \frac{\omega_0^2}{\omega_0^2 - \omega^2 - i \gamma \omega}, \\
    \mu(\omega) & = \mu_{\infty} + f_m \frac{\omega_0^2}{\omega_0^2 - \omega^2 - i \gamma \omega},
\end{split}
\end{align}
where $\omega_0$ is the resonant transition frequency, $\gamma$ is the transition decay rate, $\varepsilon_\infty$ ($\mu_\infty$) is the background permittivity (permeability), and $f_e$, $f_{me}$, and $f_m$ are the electric, magneto-electric, and magnetic oscillator strengths of the particular molecular transition \cite{baranov2023toward}. These strengths should be expressed via the microscopic parameters of the medium; for simplicity, we will relate them to transition dipole moments of the medium:
\begin{equation}
\label{eq:trdipmom}
    f_e = A |\mathbf{p}_{01}|^2, \quad
    f_{me} = A \Im [ \mathbf{p}_{01} \cdot \bm_{01}^* ], \quad
    f_m = A |\mathbf{m}_{01}|^2
\end{equation}
where $\mathbf{p}_{01}$ and $\mathbf{m}_{01}$ are the fundamental electric and magnetic transition dipole moments of the molecular resonance, and $A$ is a constant common for all three oscillator strengths. For a bi-isotropic medium (to which we limit our analysis in this work) these are related by a simple expression \cite{baranov2023toward}:
\begin{equation}
\label{eq:biiso}
    \mathbf{m}_{01} = - i \xi \mathbf{p}_{01},
\end{equation}
where $\xi$ is a real-valued microscopic chirality parameter; in the absence of higher-energy transitions ($\varepsilon_\infty = \mu_\infty = 1$), $\xi = \pm 1$ describes an ideal left-handed ($+1$) or right-handed ($-1$) chiral medium.

In the next section, we will use a numerical example to demonstrate that the pole in the macroscopic parameters $\varepsilon$, $\mu$, and $\kappa$ causes strong coupling between photonic modes of the resonator and the medium. To reveal the chirality-associated features of this effect and to describe the strong coupling analytically, below we use an approximation of small oscillation strengths. Within this approximation, the frequency-dependent part in formulas \eqref{eq:lorenz} changes macroscopic parameters from their background values only slightly, even in the vicinity of the frequency $\omega_0$.
Under these conditions, in the vicinity of the frequency $\omega_0$ one can write an approximate formula for the frequency dependence of refractive indices of the LH and RH plane waves:
\begin{align}
\label{eq:epskapapp}
\begin{split}
    \sqrt{\varepsilon\mu}&\pm \kappa \approx \sqrt{\varepsilon_\infty\mu_\infty}+\\
    &+\left(\frac{f_e}{2}\sqrt{\frac{\mu_\infty}{\varepsilon_\infty}} + \frac{f_m}{2}\sqrt{\frac{\varepsilon_\infty}{\mu_\infty}}\pm f_{me}\right)\frac{\omega_0^2}{\omega_0^2 - \omega^2-i\mathrm{\gamma}\omega}.
\end{split}
\end{align}
As follows from Eqn.~\eqref{eq:epskapapp}, in the entire spectral range of interest, the asymmetry in the interaction of the RH and LH plane waves with the chiral medium is induced by non-zero magneto-electric oscillator strength $f_{me}$. Considering only a medium with the macroscopic chirality coefficient $\kappa$ not exceeding  $\sqrt{\varepsilon\mu}$ \cite{seidov2025unbounded} \footnote{Equality $\sqrt{\varepsilon\mu} = \kappa$ can be considered as a critical condition for the medium beyond which the properties of the medium is drastically change.}, we obtain that the maximal asymmetry in the refractive indexes of LH and RH plane waves is achieved when the oscillator strength $f_{me}$ is taken as
\begin{equation}
    \label{eq:fx}
    f_{me} = \frac{1}{2}\left(\sqrt{\frac{\mu_\infty}{\varepsilon_\infty}}f_e+\sqrt{\frac{\varepsilon_\infty}{\mu_\infty}}f_m\right) \equiv f_{me}^{\mathrm{cr}}.
\end{equation}
Indeed, after substitution \eqref{eq:fx} into \eqref{eq:epskapapp}, we obtain that for such $f_{me}$ 
\begin{align}
\begin{split}
    &\sqrt{\varepsilon\mu}-\kappa = \sqrt{\varepsilon_\infty\mu_\infty}\\
    &\sqrt{\varepsilon\mu}+\kappa = \sqrt{\varepsilon_\infty\mu_\infty}
    +2f_{me}\frac{\omega_0^2}{\omega_0^2 - \omega^2-\mathrm{i}\mathrm{\gamma}\omega}
\end{split} 
\end{align}
The first case corresponds to the situation when the resonant part of polarization (and magnetization) of the medium, induced by the electric field (and magnetic field), is fully compensated by the gyrotropic contribution. This implies that in the case considered, the chiralities of light and medium are exactly opposite, which makes them non-interacting with each other. The resonant frequency under this condition can be written analytically:
\begin{equation}
    \frac{\omega}{c} = \frac{-\mathrm{arg} ~r \pm 2\alpha + \pi N}{g\sqrt{\varepsilon_\infty\mu_\infty}}.
\end{equation}
The second case describes the situation when the chiralities of the optical mode and medium match with each other, leading to strong coupling between them with the largest interaction constant. To find resonant wavelengths in this case, transcendental equation \eqref{eq:finalg} has to be solved. 

Taking into account expressions~\eqref{eq:trdipmom} and~\eqref{eq:biiso} one can obtain that the critical value of the magneto-electric oscillator strength, at which chiral light-matter interaction is maximal, is achieved at the following chirality parameter
\begin{equation}
    \label{eq:cr}
    \xi_\mathrm{cr} = -\sqrt{\frac{\mu_\infty}{\varepsilon_\infty}}.
\end{equation}
The ratio $\xi/\xi_\mathrm{cr}$ has a meaning of a normalized microscopic chirality parameter of the medium in the presence of a non-resonant background. It can be shown that within the approximation of small oscillator strengths,
\begin{equation}
\label{eq:conn}
\frac{2\xi/\xi_\mathrm{cr}}{1+\left(\xi/\xi_\mathrm{cr} \right)^2} = \frac{\kappa(\omega)}{\sqrt{\varepsilon(\omega)\mu(\omega)}-\sqrt{\varepsilon_{\infty}\mu_{\infty}\phantom{(}}}. 
\end{equation}
The parameter on the right-hand side of this formula is analogous to the macroscopic relative chirality parameter \eqref{eq:rel} with the only exception that the latter does not contain the term $\sqrt{\varepsilon_\infty\mu_\infty}$ in the denominator. Hence, the equality \eqref{eq:conn} can be interpreted as a relation between the medium's chirality parameters obtained from macroscopic and microscopic points of view.

Interestingly, the critical value of the microscopic chirality parameter \eqref{eq:cr} up to the minus sign is exactly the impedance of the non-resonant background medium $Z = \sqrt{\mu_\infty / \eps_\infty}$. This can be understood by recalling that the electric and magnetic fields of chiral plane waves propagating in the background medium are related by the same exact impedance \cite{baranov2023toward}: $Z \vec{H} = - i \Lambda \vec{E}$, where $\Lambda = \pm 1$ is the helicity of the chiral plane waves ($\Lambda=+1$ corresponds to LH plane waves, while $\Lambda=-1$ to RH plane waves). Thus, to obtain an ideal decoupling for a certain field helicity, one should engineer electric and magnetic dipole moments of the emitter such that the dipole-field interaction energy $U$ vanishes for one of the two helicities:
\begin{equation}
    U = \mathbf{p}_{01} \cdot E + \mathbf{m}_{01} \cdot H = 0,
\end{equation}
which yields the same result as \eqref{eq:cr}.

Under the condition of small oscillator strengths, one can use the approximate expression for  $\sqrt{\varepsilon\mu}+\kappa$ in the form~\eqref{eq:epskapapp}, and finally obtain an approximate formula for the Rabi splitting for an ideal LH or RH chiral medium:
\begin{equation}
    \label{eq:rabi}
    \Omega = \sqrt{\frac{4\omega_0^2A|\mathbf{p}_{01}|^2}{\varepsilon_\infty}-\gamma^2}.
\end{equation}

\begin{figure*}[t!]
\centering\includegraphics[width=0.8\textwidth]{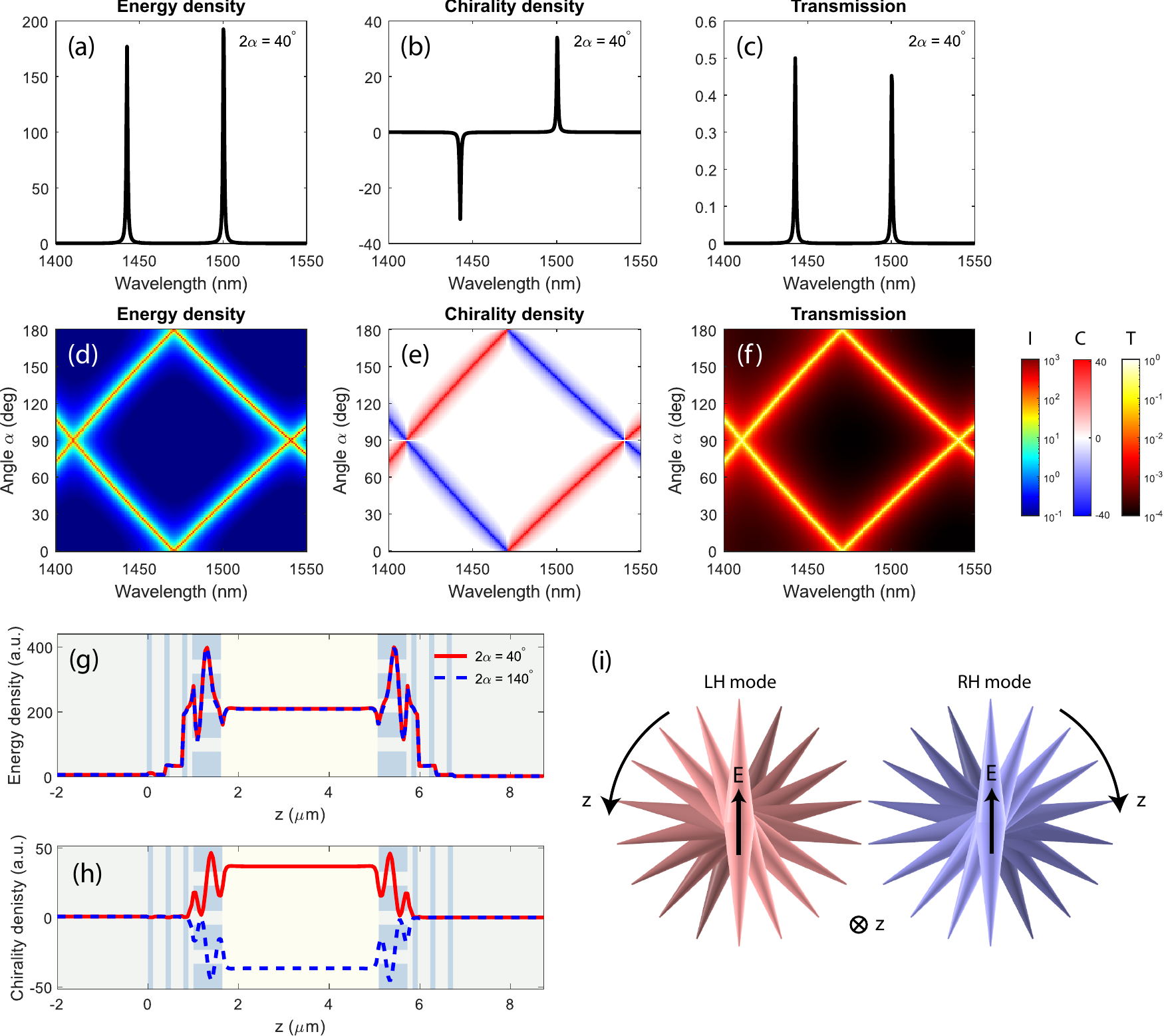}
\caption{Spectral dependencies of (a) the local field energy density $I = \varepsilon|\vec{E}|^2+\mu|\vec{H}|^2$, (b) the local chirality density $C = \varepsilon\mu\mathrm{Im}(\vec{E}\cdot\vec{H}^*)$, and (c) the transmission coefficient $T$ calculated for the twist angle $2\alpha = 40^\circ$ and a gap size $g=3450$~nm. Panels (d), (e), and (f) present these parameters as two-dimensional functions of the wavelength and the twist angle $2\alpha$. Panels (g) and (h) show the local field energy density and chirality density as a function of the $z$ coordinate for $x=y=0$, computed at $2\alpha=40^\circ$ and $140^\circ$, corresponding to modes with opposite handedness. (i) --- conceptual depiction of the instantaneous field distribution for left-handed (LH) and right-handed (RH) chiral Fabry-Pérot modes. Panels (a)--(h) are calculated for the RH polarization of incident light.}
\label{fig:fabrynonchiral}
\end{figure*}

Below, we present a numerical example to illustrate our theoretical predictions.

\section{Numerical example}
In this section, we present a specific implementation of the chiral Fabry-Pérot resonator using anisotropic mirrors that maintain handedness. A chiral medium is positioned between these mirrors, and we examine the characteristics of this configuration employing a Moiré adaptation of the Fourier modal method \cite{salakhova2021fourier}, alongside a Fourier modal method framework tailored for chiral media \cite{smagin2025}.
\subsection{Wideband handedness-preserving mirror}
\label{sec:mirror}
We start with a handedness-preserving mirror operating over the frequency range with a width significantly exceeding those of the Fabry-Pérot modes and the material resonance of the cavity medium. As mentioned in Sec.~\ref{sec:theory}, an anisotropic mirror preserves the handedness of normally incident light if the amplitude reflection coefficients of two orthogonal polarizations are opposite, i.e., $r_{xx} = -r_{yy}$. If, in addition to this, the absolute values of the amplitude reflection coefficients are close to unity, $|r_{xx}|^2=|r_{yy}|^2=1$, the Fabry-Pérot resonator comprising a pair of such mirrors will exhibit modes with a high quality factor. To satisfy both requirements, we construct a mirror from a one-dimensional array of silicon stripes in a silica matrix on a Si/SiO$_2$ multilayered structure. The array of stripes introduces in-plane anisotropy into the system, enabling the condition of the anti-phase amplitude reflection to be satisfied, while the multilayered structure ensures high reflection coefficients across the wide frequency range.

To identify geometric parameters of the structure that satisfy these requirements, we optimize the layers' thicknesses, the grating period, and its filling ratio using a genetic algorithm. In optimization, we assume that the dielectric permittivity and magnetic permeability of the incoming medium are constant and equal to $\varepsilon = 2$ and $\mu = 1$. The optimal geometric parameters of the mirror are provided in Fig.~\ref{fig:mirror}. In principle, replacing the array of stripes with a homogeneous layer of anisotropic material exhibiting in-plane anisotropy could be a viable alternative. However, such a configuration requires anisotropic material with rather large in-plane anisotropy, which could be difficult to achieve in practice. See the Supplemental Materials for an example of a wideband handedness-preserving mirror based on a homogeneous layer of anisotropic material with unrealistically large anisotropy. 

\begin{figure*}[t!]
\centering\includegraphics[width=1\textwidth]{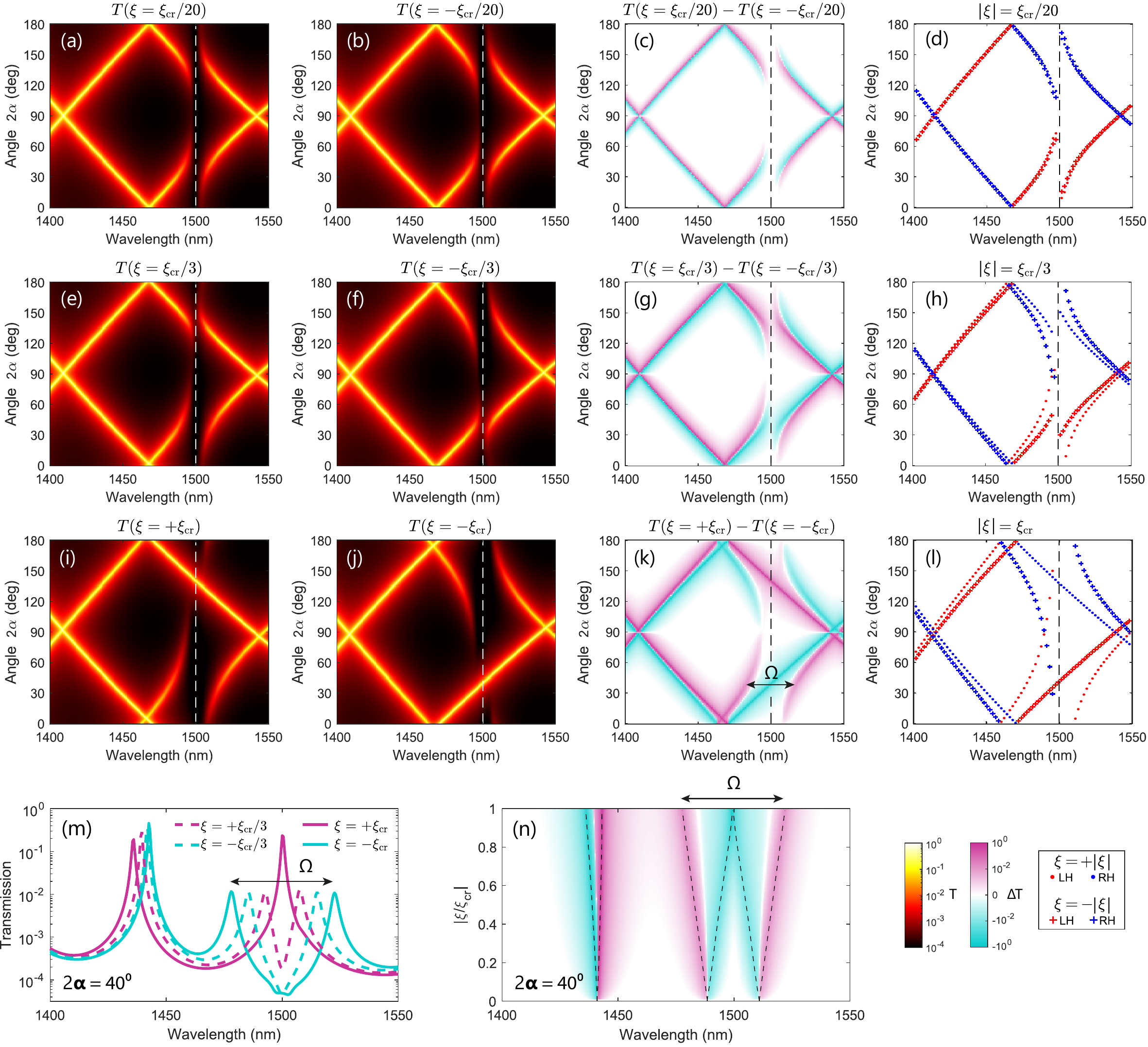}
\caption{(a,b,e,f,i,j) Wavelength and twist-angle dependencies of the resonator transmission coefficient and (c,g,k) differential transmission coefficient for different microscopic chirality parameters $\xi$, representing small, intermediate and maximal material chirality. (d,h,l) Spectral positions of the resonances calculated by formula \eqref{eq:finalg}. Red and blue colors correspond to LH and RH modes. Dots and crosses correspond to positive and negative microscopic chiralities. Vertical dashed lines in panels (a--l) denote the spectral position of the pole in material parameters $\varepsilon$, $\mu$ and $\chi$. (m) Transmission spectra of the resonator calculated for intermediate and maximal material chiralities. (n) Differential transmission spectrum calculated as a function of the material chirality parameter $\xi$ varying between 0 and 1. Dashed lines in (k) represent local extrema in differential transmission. All panels are calculated for the RH polarization of incident light. Parameters of the Lorentz model are the following: $\varepsilon_\infty =2$, $\mu_\infty = 1$, $2\pi c/\omega_0 = 1500$~nm, $\gamma_0 = \omega_0/300$, $f_e\omega_0^2 = 500$~meV$^2$, $f_m = \xi^2f_e$, $f_{me} = -\xi f_e$.}
\label{fig:strongcoupling}
\end{figure*}

As illustrated in Fig.~\ref{fig:mirror}b,c, the conditions $r_{xx} = -r_{yy}$ and $|r_{xx}|^2 = |r_{yy}|^2 = 1$ are almost met for a designed mirror over the wavelengths between 1400 and 1800~nm, that ensures the high quality factor of the Fabry-Pérot modes in this spectral range. Besides that, this mirror remains handedness-preserving even in the wide range of incident angles (See Supplemental materials for the spectral-angular dependencies of cross-polarized reflection coefficients.)


\subsection{Chiral Fabry-Pérot modes}
Next, we conduct an analysis of a Fabry-Pérot resonator comprising a pair of wideband handedness-preserving mirrors, designed in the previous section, separated by a \textit{non-chiral} dispersionless medium with $\varepsilon = 2$ and $\mu = 1$. To study the properties of such a system, we subject the resonator to irradiation by a normally incident plane wave coming in from the far field and traveling in the positive $z$-direction. Fig.~\ref{fig:fabrynonchiral}(a)--(c) shows the spectral dependencies of the local field energy density, $I$, local chirality density, $C$, at the center of the cavity, as well as the transmission coefficient, $T$. The local field energy density and the chirality density are calculated as
\begin{align}
I & = \varepsilon|\vec{E}|^2+\mu|\vec{H}|^2\\
C & = \varepsilon\mu\cdot\mathrm{Im}(\vec{E}\cdot\vec{H}^*)
\end{align}
Calculations are made for the gap size between the mirrors $g = 3450$~nm, the twist angle $2\alpha=40^\circ$, and the in-plane wavevector $k_x = k_y = 0$. It can be seen from Fig.~\ref{fig:fabrynonchiral}(a)--(c) that all spectra have sharp peaks. The wavelengths of the peaks 1443~nm and 1500~nm are well described by formula~\eqref{eq:finalg} by taking $N = 7$ and $\mathrm{arg}~r \approx 1.44$ for the first resonance, which corresponds to the RH wave, and $N = 7$ and $\mathrm{arg}~r \approx0.86$ for the second resonance, which corresponds to the LH wave. It proves that the peaks are indeed the Fabry-Pérot resonances of the system. 
The mentioned values of $\mathrm{arg} r$ correspond to the amplitude reflection coefficients of the mirrors calculated in the previous section using the Fourier modal method. The peak values of local energy density for both resonances falling within the calculation range are almost the same, while their chirality densities are opposite. Due to the high efficiency of handedness-preserving mirrors within the entire range of interest, both modes have a high quality factor and a high local field enhancement. It ensures that the modes are efficiently excited by any polarization state of the incoming wave (See supplemental material for details). From a practical viewpoint, transmission spectra are important because they enable us to probe the spectral position of the structure's eigenmodes by collecting the field signal. Note that although the transmission spectra also possess sharp peaks indicating the existence of the modes, the peak's position and intensity do not contain any information about the handedness of the modes.

The effect of the twist angle variation on the spectra of parameters $I$, $C$, and $T$ is demonstrated in Fig.~\ref{fig:fabrynonchiral}(d)--(f). It is evident that with an increase in the twist angle from 0 to $\pi$, the spectral positions of the modes change in correspondence with formula~\eqref{eq:finalg}. When $2\alpha = 0, \pi/2$, and $\pi$, the modes are degenerate, as predicted in Sec.~\ref{sec:theory:FP}. At these angles, the chirality density of the field in the cavity is exactly zero, which can be seen by the white crosses in Fig.~\ref{fig:fabrynonchiral}(e). Fig.~\ref{fig:fabrynonchiral} also reveals a remarkable property of the Fabry-Pérot resonator under study: for any wavelength of interest, the twist angle between the mirrors can be adjusted to satisfy the resonance condition. Moreover, this property is a part of an even stronger statement: the resonance condition can be satisfied for any fixed parameter among the gap size, refractive index of chiral medium, wavelength, and twist angle by appropriately adjusting the remaining two parameters. The effect of gap size variation is studied for a similar structure in Ref.~\cite{dyakov2024chiral}. This tunability makes the proposed structure prospective for the practical realization of the enhanced chiral light-matter interaction.

So far, we have identified the handedness of a Fabry-Pérot mode based on the sign of the chirality density or on whether the eigenvector to which it corresponds is $[1, 0]^T$ or $[0, 1]^T$. Let us now explicitly demonstrate that, indeed, the field's spatial distribution of the twisted Fabry-Pérot resonator cannot be superimposed on its mirror image form. For this, in Fig.~\ref{fig:fabrynonchiral}(i) we plot the instantaneous electric vector field excited by the incoming wave within the gap region. The field distribution inside the gap is a result of the incident wave being refracted into the gap region and then subsequently multireflected from the upper and lower mirrors with the preservation of handedness. As a result of the interference between two counter-propagating circularly-polarizing waves of the same handedness, a standing wave is formed. As can be seen from Fig.~\ref{fig:fabrynonchiral}(i), locally, this standing wave is linearly polarized; however, the polarization direction forms a helix with the z-coordinate. Despite being locally linearly polarized, such a standing wave obviously cannot be superimposed on its mirror image and, thus, is chiral.

The $z$-profiles of the time-average local field intensity and chirality density settled within the structure by the incoming wave are shown in Fig.~\ref{fig:fabrynonchiral}(g,h). Although the field in the gap region is a result of the interference between two counter-propagating waves, both parameters $I$ and $C$ are homogeneous in the gap region between the mirrors (except for the regions in the near-field of the mirrors). As shown in Ref.~\cite{dyakov2024chiral}, the absence of the nodes and antinodes in a standing wave is a result of the circular polarization of interfering waves and a specific requirement for the mirrors to satisfy the condition $r_{xx}=-r_{yy}$. 


\subsection{Strong coupling in cavity of the Fabry-Pérot resonator}
In the previous section, we have described chiral photonic modes supported by the twisted Fabry-Pérot resonator with a non-chiral medium between the mirrors. Let us now add \textit{chiral} matter into this system, placing a material with a non-zero macroscopic chirality parameter $\kappa$ into the gap region of the resonator. We will use a material with the Lorentz pole in the macroscopic parameters $\varepsilon(\omega)$, $\mu(\omega)$, and $\kappa(\omega)$ set by formulas \eqref{eq:lorenz}. To study the interaction of the photonic mode with matter in such a system, we calculate the transmission coefficient of a normally incident plane wave as a function of the wavelength and the twist angle between the mirrors, in the same range of these variables as in Fig.~\ref{fig:fabrynonchiral}(f). 

Fig.~\ref{fig:strongcoupling} shows the set of transmission maps calculated for LH and RH enantiomers of the chiral medium, and for a small, intermediate, and maximal microscopic chirality parameter. In the case of the small microscopic chirality parameter ($\xi = \xi_\mathrm{cr}/20$), i.e. weak chirality of the medium [see Fig.~\ref{fig:strongcoupling}(a,b)], the transmission maps for the left and right medium's enantiomers have two avoided intersections of the material pole (vertical white dashed lines) with photonic modes. Avoided intersections suggest that the interaction of the material pole with the photonic mode is governed by a strong coupling mechanism. The difference between the two maps shown in Fig.\ref{fig:strongcoupling}c indicates that the modes' positions on the two maps are only slightly different, which is due to the slightly different light-matter coupling conditions in terms of matching between the types of light and matter enantiomers. 

For the intermediate chirality of the medium ($\xi = \xi_\mathrm{cr}/3$), the upper and lower avoided intersections on each map are already notably different [see Fig.~\ref{fig:strongcoupling}(d,e)]. This is attributed to even more distinct coupling strengths between light and matter of different handednesses. Note that, like in the case of a weak medium chirality, the maps for the left and right media are mirror-symmetrical to each other with respect to the line $2\alpha = 90^\circ$. This is because the properties of the Fabry-Pérot resonator with a twisted angle $2\alpha$ are obviously inverted (in terms of handedness of the photonic modes) with respect to those with a twist angle of $180-2\alpha$.

In the case of the maximal medium chirality ($\xi = \xi_\mathrm{cr}$), one of the avoided intersections on each of the maps (Fig.~\ref{fig:strongcoupling}(g,h)) disappears. It suggests that one of the two photonic modes no longer interacts with the medium, while the other one interacts with the medium with the largest interaction constant, in full correspondence with our prediction in Sec.~\ref{sec:theory:FP}. The Rabi splitting determined from Fig.~\ref{fig:strongcoupling}(f) $(2\pi c/\Omega)_{\mathrm{sim}}\approx 45$~nm is well described by formula \eqref{eq:rabi} $(2\pi c/\Omega)_{\mathrm{theor}}\approx 49$~nm. Note that the decoupling takes place for opposite enantiomers of the medium and photonic modes with different signs of the parameters $\xi$ and $\Lambda$).

The difference in how the LH and RH chiral media interact with a photonic mode of a particular handedness results in distinct transmission spectra for the Fabry-Pérot resonator. Figure~\ref{fig:strongcoupling}(j) displays the calculated transmission spectra at a twist angle of $2\alpha= 40^\circ$, comparing scenarios of intermediate medium chirality (represented by dashed lines) and maximal medium chirality (depicted by solid lines). The figure illustrates that the spectra for the left and right media differ in both instances, with the most significant variation occurring near the material pole $\omega_0$. For the maximal medium chirality, the transmission spectrum, corresponding to the opposite handednesses of the light and matter, reveals a single peak near the pole, representing a photonic mode of the resonator that remains unaffected by its interaction with the medium. 

Finally, Fig.~\ref{fig:strongcoupling}(k) shows the difference between transmission spectra for the left and right media for the medium's normalized microscopic chirality parameter $\xi/\xi_\mathrm{cr}$, continuously varying between 0 and 1. One can see that as the chirality of the medium increases, the peaks nearest to the transition frequency $\omega_0$ progressively converge towards each other, culminating in a unified peak at $\xi = \xi_\mathrm{cr}$.

\section*{Conclusion} 
In summary, we have studied the interaction of chiral electromagnetic modes with chiral matter within a macroscopic electrodynamics framework. To obtain chiral electromagnetic modes, we have developed a wideband handedness-preserving mirror based on one-dimensional periodic gratings placed on top of the Bragg mirror. Using a pair of these mirrors, we have formed a Fabry-Pérot resonator that supports chiral modes with eigenfrequencies dependent on the angle between the orientation of the upper and lower gratings. To model the chiral medium, we have used an approximation of the Pasteur medium, which is characterized by the local macroscopic chirality parameter $\kappa$. To properly describe the interaction of chiral modes with the chiral medium, we have introduced the Lorentz pole into the macroscopic parameters $\varepsilon(\omega)$, $\kappa(\omega)$, and $\mu(\omega)$ with oscillator strengths connected via the microscopic chirality parameter $\xi$. Then, we have placed the Pasteur medium into the Fabry-Pérot resonator between the mirrors. Using the formalism of Jones matrices, we have demonstrated that material resonance in the medium's macroscopic parameters induces strong coupling between chiral Fabry-Pérot modes and the chiral medium, with the Rabi splitting governed by the enantiomeric compatibility of the mode and the medium. Finally, we have confirmed our theoretical findings by comparing them with the results of full-wave simulations made by the Fourier modal method adapted for chiral media. 

We believe that our work provides a foundation for future research on chiral light-matter interactions, paving the way for innovative optical sensing of chiral organic molecules and enantiomer-selective technologies.

\section*{Acknowledgement}
This work was supported by the Russian Science Foundation (Grant No. 25-12-00454).
%

\end{document}